
\input harvmac.tex
%
%
%
%
\ifx\answ\bigans
\else
\output={
  \almostshipout{\leftline{\vbox{\pagebody\makefootline}}}\advancepageno
}
\fi
%
%
%
\def\mayer{\vbox{\sl\centerline{Department of Physics 0319}%
\centerline{University of California, San Diego}
\centerline{9500 Gilman Drive}
\centerline{La Jolla, CA 92093-0319}}}
%
%

%
%
\def\UCSD#1#2{\noindent#1\hfill #2%
\bigskip\supereject\global\hsize=\hsbody%
\footline={\hss\tenrm\folio\hss}}
%
%
\def\abstract#1{\centerline{\bf Abstract}\nobreak\medskip\nobreak\par #1}
%
%
%
%
\edef\tfontsize{ scaled\magstep3}
 \tfontsize  \tfontsize
 \tfontsize \font\titlei=cmmi10 \tfontsize
\font\titleis=cmmi7 \tfontsize \font\titleiss=cmmi5 \tfontsize
\font\titlesy=cmsy10 \tfontsize \font\titlesys=cmsy7 \tfontsize
\font\titlesyss=cmsy5 \tfontsize  \tfontsize
\skewchar\titlei='177 \skewchar\titleis='177 \skewchar\titleiss='177
\skewchar\titlesy='60 \skewchar\titlesys='60 \skewchar\titlesyss='60
%
%
%
%
%
\def\inv{^{\raise.15ex\hbox{${\scriptscriptstyle -}$}\kern-.05em 1}}
\def\lbar{{\lower.35ex\hbox{$\mathchar'26$}\mkern-10mu\lambda}} 

%
%
%
%
\def\dsl{\,\raise.15ex\hbox{/}\mkern-13.5mu D} 
\def\delsl{\raise.15ex\hbox{/}\kern-.57em\partial}
\def\Ksl{\hbox{/\kern-.6000em\rm K}}
\def\Asl{\hbox{/\kern-.6500em \rm A}}
\def\Dsl{\hbox{/\kern-.6000em\rm D}} 
\def\Qsl{\hbox{/\kern-.6000em\rm Q}}
\def\gradsl{\hbox{/\kern-.6500em$\nabla$}}
%
%
\def\lspace{\ifx\answ\bigans{}\else\qquad\fi}
\def\lbspace{\ifx\answ\bigans{}\else\hskip-.2in\fi} 
%
%
\def\boxeqn#1{\vcenter{\vbox{\hrule\hbox{\vrule\kern3pt\vbox{\kern3pt
        \hbox{${\displaystyle #1}$}\kern3pt}\kern3pt\vrule}\hrule}}}
%
%
\def\mbox#1#2{\vcenter{\hrule \hbox{\vrule height#2in
\kern#1in \vrule} \hrule}}
%
%
%
%

  \def\CO{{\cal O}}

%
%
%
%
%

%

%
%

\def\darr#1{\raise1.5ex\hbox{$\leftrightarrow$}\mkern-16.5mu #1}

%
%
\def\half{{\textstyle{1\over2}}} 
\def\frac#1#2{{\textstyle{#1\over #2}}} 
%
%
%
%

%
%
%
%

%
%
\def\ltap{\ \raise.3ex\hbox{$<$\kern-.75em\lower1ex\hbox{$\sim$}}\ }
\def\gtap{\ \raise.3ex\hbox{$>$\kern-.75em\lower1ex\hbox{$\sim$}}\ }
\def\gl{\ \raise.5ex\hbox{$>$}\kern-.8em\lower.5ex\hbox{$<$}\ }
\def\roughly#1{\raise.3ex\hbox{$#1$\kern-.75em\lower1ex\hbox{$\sim$}}}
%
%

%

%
\def\np#1#2#3{{Nucl. Phys. } B{#1} (#2) #3}
\def\pl#1#2#3{{Phys. Lett. } {#1}B (#2) #3}
\def\prl#1#2#3{{Phys. Rev. Lett. } {#1} (#2) #3}
\def\physrev#1#2#3{{Phys. Rev. } {#1} (#2) #3}

\relax

\def\CO{{\cal O}}

\def\lta{\ \hbox{\raise.55ex\hbox{$<$}} \!\!\!\!\!
\hbox{\raise-.5ex\hbox{$\sim$}}\ }
\def\gta{\ \hbox{\raise.55ex\hbox{$>$}} \!\!\!\!\!
\hbox{\raise-.5ex\hbox{$\sim$}}\ }
\def\mayer{\vbox{\sl\centerline{Department of Physics}
\centerline{9500 Gilman Drive 0319}
\centerline{University of California, San Diego}
\centerline{La Jolla, CA 92093-0319}}}

\def\frac#1#2{{\textstyle{#1 \over #2}}}

\def\[{\left[}
\def\]{\right]}
\def\({\left(}
\def\){\right)}
\def\lfm{\medskip\noindent}
\noblackbox
\vskip 1.in

\centerline{{\titlefont{Debye
screening    and baryogenesis}}}
\vskip .15in
\centerline{{\titlefont{  during the  electroweak phase
transition}}}
\medskip
\centerline{{\bf Submitted to Physics Letters B}}
\bigskip
\medskip
\centerline{A. G. Cohen\footnote{}{Email:
cohen@andy.bu.edu, dkaplan@ucsd.edu, anelson@ucsd.edu}
\footnote{$^a$}{DOE Outstanding Junior Investigator}}
\centerline{{\sl
Physics Department}} \centerline{{\sl Boston University}}
\centerline{{\sl Boston, MA 02215}}
 \medskip
\centerline{ D. B. Kaplan$^a$\footnote{$^b$}{
Sloan Fellow}\footnote{$^c$}{NSF Presidential Young Investigator}  and A. E.
Nelson $^b$\footnote{$^d$}{SSC fellow}
  }
\bigskip\mayer \bigskip \vfill
\abstract{We examine a recent claim that Debye screening will affect
the charge transport mechanism of anomalous electroweak  baryogenesis, and
show that the  effects of gauge charge screening do not affect the baryon
number produced during a first order electroweak phase transition.  }
\vfill\UCSD{\hbox{UCSD/PTH 92-19,\ \break BU-HEP-92-20}}{June 1992}

 The realization that the
electroweak anomalous baryon violation is probably rapid enough to be
cosmologically relevant at high temperatures \nref\early{A. Linde,
\pl{70}{1977}{306}; S. Dimopoulos and L. Susskind, \physrev{D18}{1978}{4500};
 N. Christ, Phys. Rev. D21 (1980) 1591; N.S. Manton, Phys.
Rev. D28 (1983) 2019; F.R. Klinkhammer and N.S. Manton, Phys. Rev. D30
(1984) 2212}\nref\krs{V.A. Kuzmin, V.A.
Rubakov and M.E. Shaposhnikov, \pl{155}{1985}{36} }\nref\more{P. Arnold and L.
McLerran,   Phys. Rev. D36 (1987) 581;
   Phys. Rev. D37 (1988)
1020;  S. Khlebnikov and M. Shaposhnikov, \np{308}{1988}{885};  M. Dine, O.
Lechtenfeld, B. Sakita, W. Fischler and J. Polchinski, Nucl. Phys. B342 (1990)
381
}\refs{\early-\more} has led to several new proposals for  weak scale
baryogenesis  \nref\shaposh{M.E. Shaposhnikov, JETP Lett. 44 (1986) 465;
\np{287}{1987}{757}; \np{299}{1988}{797};
 A. I. Bochkarev, S. Yu. Khlebnikov and M.E. Shaposhnikov,
\np{329}{1990}{493} }\nref\mclerran{ L. McLerran,
\prl{62}{1989}{1075}}\nref\usone{A.G. Cohen, D.B. Kaplan and A.E. Nelson,
\pl{245}{1990}{561};
\np{349}{1991}{727}}\nref\mtsv{N. Turok and  J. Zadrozny,
 \prl{65}{1990}{2331}; \np{358}{1991}{471}; L. McLerran, M. Shaposhnikov,
N. Turok and M. Voloshin, \pl{256}{1991}{451}}\nref\dhss{M. Dine, P. Huet,
R. Singleton and L. Susskind,
\pl{257}{1991}{351}}\nref\ustwo{A.G. Cohen, D.B. Kaplan and A.E.
Nelson, \pl{263}{1991}{86}}\nref\usthree{A.E. Nelson, D.B. Kaplan and A.G.
Cohen, \np{373}{1992}{453}}\nref\shaptwo{ M.E. Shaposhnikov,
Phys. Lett. 277B (1992) 324  }\refs{\krs,\shaposh-\shaptwo} which may be
testable at the SSC.  In general, baryogenesis requires out of equilibrium CP
violation as well as baryon number violation  \ref\barcon{A.D. Sakharov, JETP
Lett. 6 (1967) 24}.  Departure from thermal  equilibrium can result  from a
first order weak phase transition, occurring either in the standard model or
in one of its extensions, which proceeds via nucleation and growth of bubbles
of the Higgs phase. Many extensions of the standard model can contain
sufficient CP violation to plausibly produce a baryon density
to photon entropy ratio  as
large as the observed $\rho_B/s\sim 10^{-10}$ \refs{\usone-\usthree}.
The  mechanism through which weak scale
baryogenesis can proceed depends on the thickness of the bubble walls
produced during the phase transition, compared to relevant mean free paths
\foot{Recent work on the standard model \ref\smw{M. Dine, R.G. Leigh, P. Huet,
A. Linde and D. Linde, Slac preprint SLAC-PUB-5741 (1992); B.H. Liu, L.
McLerran, and N. Turok, Minnesota preprint TPI-MINN-92/18-T (1992).}\ and
minimal
supersymmetric standard model \nref\mssm{G.F. Guidice, Texas preprint
UTTG-35-91 (1991); S. Myint, Boston University preprint BU-HEP-92-4
(1992)}\nref\usfour{A.G. Cohen,  D.B. Kaplan and A.E. Nelson, work in progress
}\refs{\mssm,\usfour} suggests that the bubble walls there are thick, ($\sim
(25-40)/T$).  Baryogenesis arising from the singlet majoron model \usone\ can
involve thin bubble walls ($\sim 1/T $) \ref\smaj{The wall thickness in the
singlet majoron model for some parameters can be simply computed from the
results
of Y. Kondo, I. Umemura, and K. Yamamoto, \pl{263}{1991}{86}}, as can generic
models with several scalars.}.  A generic mechanism for thick walls is
spontaneous baryogenesis, first
introduced to explain  baryogenesis during a second order phase transition
\ref\ck{A. Cohen and D.B. Kaplan, \pl{199}{1988}{251}; \np{308}{1988}{913}}.
Spontaneous baryogenesis involves coherent effects of a time and space
dependent
scalar field acting as a charge potential. The baryon number is produced
inside the bubble walls  where  scalar field expectation values (like
the Higgs field vev)
are changing, causing  the  free
energy density inside the bubble walls to be minimized  for  nonzero baryon
number \refs{\mtsv-\ustwo}.  However the anomalous baryon number
violating processes have a rate per unit volume
 which, even at high temperatures,  is rather slow, and
$\rho_B/s$ produced inside the walls turns out to be {\it at most} of order
$(\alpha_{\rm wk}^4/g_*)\delta_{CP}\sim 10^{-8}\delta_{CP}$. Here
$\delta_{CP}$ is a reparametrization invariant CP violating phase and
$g_*=\CO(100)$ is the effective number of particle degrees of freedom at the
electroweak transition
\nref\dhs{M. Dine, P. Huet, and R. Singleton, UCSC preprint
SCIPP-91-08 (1991), to appear in Nuclear Physics B}\nref\gst{D. Grigoriev, M.
Shaposhnikov, and N. Turok, Princeton preprint PUPT-91-1274
(1991)}\refs{\mtsv-\ustwo,\dhs,\gst}.

 We have proposed in refs.~\refs{\usone,\usthree} a   more efficient
``charge transport mechanism'' for producing   baryon number   during
the electroweak  phase transition in models with thin bubble walls.
In a  CP violating theory,  the reflection probabilities from the phase
boundary for   left handed weak doublets and their CP conjugates are
different, and so the bubble wall  reflects  a CP asymmetric distribution of
particles into the unbroken phase.    Anomalous baryon violating electroweak
processes throughout a large region of the unbroken phase will  be biased
toward producing baryons wherever there is net lefthanded fermion number
(i.e. where there  are more righthanded antiquarks and antileptons than
lefthanded quarks and leptons). This mechanism dominates over mechanisms which
produce baryons only inside the bubble walls unless the walls are   thicker
than the relevant particle mean free paths (which are about
$(1-10)/T $), because the anomalous baryon number violating processes
are biased over a larger region.   In ref.~\usthree\ we simulated the fermion
reflection and scattering processes and found that a significant asymmetry in
fermion quantum numbers extends in front of the wall over a volume
considerably larger than the volume of the walls. We then  showed  that
baryon to entropy ratios as large as $\sim 10^{-5}\delta_{CP}$ are possible
in a two Higgs model if the thickness of the walls is of order the inverse
top mass, and the walls are moving with velocities of order $0.1 c$.
 However the charge density in the unbroken phase that we computed included a
net hypercharge,  and so one
should consider the effects of Debye screening on the particle number
distributions, and on baryogenesis\nref\deb{L.
McLerran, private communication; M. Dine, private
communication; A. Dolgov, private communication}\nref\khlebnikov{S. Yu.
Khlebnikov, UCLA preprint UCLA/92/TEP/14 (1992)} \refs{\deb,\khlebnikov}.
In a recent quantitative analysis
Khlebnikov has shown that it is not possible for a net  gauge charge to
penetrate   into the unbroken phase further than the Debye screening
length,  unless the bubbles are expanding ultra-relativistically \khlebnikov.
In this letter we examine the effects of hypercharge  screening on
baryogenesis and show that this
 screening has no effect on the baryon number produced during the
transition, even in the limit of zero Debye length.

 We wish to reconsider the mechanism of ref.~\usthree\ for
baryogenesis in a first order weak phase transition. We showed that
when bubbles of the broken phase nucleate and expand into the unbroken
phase plasma, CP violation   leads to different  reflection probabilities for
lefthanded fermions and their CP conjugates  from the bubble wall. This
CP violating effect can be large for     top quarks in the two Higgs doublet
model, or for higgsinos in   supersymmetric models \refs{\usthree,\usfour}.
Thus
the bubble walls reflect nonzero values for some quantum numbers into the
unbroken phase. Furthermore, as we showed, the quantum numbers are typically
carried by particles with large momenta, and numerical simulations find that
the high momenta reflected particles penetrate a long distance,
$ \CO(10-1000)/T $, into the unbroken phase before being
rethermalized  by particle scattering. Thus a CP violating asymmetry in
particle quantum numbers extends over a large region in front of the
advancing bubble wall, unless the walls are ultra-relativistic. The particle
distribution functions in front of the wall will approach local thermal and
chemical equilibrium, characterized by a temperature and by chemical
potentials for the various conserved or approximately conserved charges.
Because left and right handed particles carry different values of weak
hypercharge, which is a conserved quantum number in the unbroken phase, we
denoted the quantum number  being reflected off the bubble wall as
hypercharge, and noted that in the presence of nonzero hypercharge the free
energy of the unbroken phase is minimized for nonzero baryon number.
Therefore anomalous electroweak baryon violating processes in the unbroken
phase will   lower the free energy by producing the baryons we see today,
which eventually enter the  bubbles of broken phase (where electroweak baryon
number violation is extremely suppressed).

A potential problem for this baryogenesis mechanism is that weak
hypercharge is   gauged, and any long range gauge fields will be screened by
the plasma. Khlebnikov \khlebnikov\ has solved the Vlasov equations for the
particle distribution functions in the unbroken phase and finds  that
hypercharge can only extend over a distance of about $ 2/T $ in front of the
bubble wall, whereas our simulation of the two Higgs doublet model (which
only included quark scattering by gluon exchange, and neglected the tiny
hypercharge field strength) found   an asymmetry in the distribution functions
between left handed top quarks and right handed top antiquarks over much larger
distances. Top quark scattering with Higgs scalars does not conserve chirality,
and hence  will change the asymmetry in axial top quarks. However since all the
scattering processes conserve several gauge and global charges,  in general  an
asymmetry in left handed particle number persists   when all scattering
processes are in equilibrium.  Screening of gauge charges is also important;
the
particle distribution functions will adjust themselves to cancel the net
hypercharge carried by the top quarks outside the Debye length. Therefore one
should worry that the calculation of the baryon number in ref.~\usthree, which
found a baryon production rate proportional to the net hypercharge in the
plasma, is incorrect. Here we will show that while gauge screening does affect
the fermion distribution functions, it has no effect on the total
lefthanded fermion asymmetry, and hence no effect on baryogenesis.

Calculation of the  baryon production rate becomes simple if
the system is near thermal and chemical equilibrium.
To compute the baryon production all one needs
 is $\Gamma$,
the total rate for anomalous $B$-violating  events per unit
volume---parametrized as $\Gamma= \kappa \alpha_{\rm
wk}^4 T^4$---and $\Delta f$, the  change in the free energy  per
anomalous event. The parameter $\kappa$ has been estimated to be of order
$0.1-1.0$ \ref\estkappa{J. Ambjorn, T. Askgaard, H.
Porter and M. Shaposhnikov, Phys. Lett. 244B (1990) 479 }, while
$\Delta f$ is \eqn\freechange{
 \Delta f=\sum_{a}\Delta q_a \mu_a\ .}
Here $a$ runs over all conserved or approximately conserved charges, $\Delta
q_a$ is the change in the $a^{\rm th}$ charge per anomalous event, and
$\mu_a$ is the chemical potential for the charge $a$. Note that
many other quantum numbers
besides hypercharge are conserved or approximately conserved in the unbroken
phase, such as baryon and lepton number.
 ``Approximately conserved'' means that
the time scale for processes which violate these quantum numbers is long when
compared with the typical transport time $\tau_T$ that a reflected
particle spends in the
unbroken phase before the bubble wall catches up with it. This time was
found to be between $10/T$ and $10^4/T$ in ref.~\usthree. Since
anomalous events are slow ($\Gamma\ll T^3/\tau_T$) and the Yukawa couplings of
light particles are small,  all quantum numbers which are conserved by
perturbative gauge interactions, scalar self couplings, and the top quark
Yukawa
interactions may be taken to be approximately conserved. To compute the rate of
change of the baryon density, note that
 \eqn\leftcharge{\Delta f=\sum_{a}\Delta q_a
\mu_a = \half\sum_i \mu_i=\half\sum_i {6\over T^2}n_i=  {3\over T^2}  n_L\ .}
Here $i$ runs over all fermions which are in weak doublets, $\mu_i$ is the
chemical
potential for the $i^{\rm th}$ particle, $n_i$ is the  number density of the
$i^{\rm th}$ particle, and $n_L$ is the net density  of left handed weak
doublet
fermions. The factor of $\half$ is present because a given event only involves
one member of each weak doublet. The total rate of baryon creation per unit
volume $\dot\rho_B$ in the unbroken phase is then just \eqn\barcreate{
\dot\rho_B=-{\Delta f\over T} 3 \Gamma\ =-3 \kappa \alpha_{\rm wk}^4
T^3\sum_{a}\Delta q_a \mu_a\ , } where the factor of 3 accounts for the fact
that the  baryon number changes by 3 units per event. Our ignorance of the
details of these anomalous baryon violating processes is subsumed into the
parameter $\kappa$; given this parameter,  calculation of the baryon production
rate reduces to a calculation of the chemical potentials  $\mu_a$ for the
various
global and gauge charges.

The chemical potentials are calculated in a standard way by solving the
linear equations
 \eqn\numbdens{\rho_b=\sum_jq^b_jn_j=\sum_{j,a}q_j^b{T^2\over
6\ }{k_j}q_j^a\mu_a\ ,} where $\rho_b$ is the (fixed) charge density of the
$b^{\rm th}$ conserved charge, the sum on $j$ runs over all particle species,
$q_j^a$ is the value of the charge $a$ carried by the $j^{\rm th}$ particle,
and $k_j$ is a statistical factor with $k_j= 1$ for every Weyl fermion
and $k_j=2$ for every charged boson which is much lighter than the
temperature. For simplicity we choose a basis for the conserved charges
such that \eqn\ortho{ \sum_{j}{k_j}q_j^a q_j^b\propto\delta_{ab}\ ,}
(which we refer to as an orthogonal basis). With this basis, the number
densities of the various charges will   simply be proportional to their
chemical potentials, which would not be true in a
non-orthogonal basis.

In this
basis, only the chemical potentials associated with gauge charges are
affected by Debye screening.  Since gauge charges are never   anomalous,
$\Delta q_a$ =0 whenever $a$ is a gauge charge, which  guarantees that
$\dot\rho_B$ in eq. \barcreate\ is unaffected by the value of $\mu_Y$.
This
is the essence of our argument that gauge charge screening does not
affect baryogenesis.

This result may seem paradoxical in light of  ref.~\usthree\ where
it was shown that in
several models where net hypercharge is reflected into the unbroken phase the
baryon production rate is proportional to the hypercharge density. The
resolution of the paradox is simple: ref.~\usthree\ did not use an orthogonal
basis for the various approximately conserved charges. In an orthogonal
basis,   several conserved global quantum numbers which are orthogonal to
hypercharge are reflected into the unbroken phase along with the net
hypercharge. For instance, although no net baryon number is reflected  (to
lowest order in gauge coupling constants) there is a nonzero reflection of
the quantum number $B'=B-xY$, where $x$ is a constant chosen such that $B'$ is
orthogonal to $Y$. For example, in the three family standard model with $n$
scalar weak doublets,
\eqn\ccond{ x={\sum_{j}{k_j}q_j^Y q_j^B\over
\sum_{j}{k_j}(q_j^Y)^2} ={2\over  10+n}\ .}
 Furthermore $B'$ is changed by anomalous electroweak
processes, and in the presence of nonzero $B'$ there will be a bias toward
producing net baryon number. The buildup of net hypercharge would produce a
long range field, causing particles to move around in order to  screen it; but
the screening does  not affect $B'$ since $B'$ is orthogonal to gauge charges.

In summary, the computation of the baryon production rate in the unbroken phase
can be done as follows:
\lfm
\item{1)} Find a basis where all approximately conserved global charges  are
mutually orthogonal, and orthogonal to all gauge charges.  In this basis
the Debye screening of gauge charges does not affect the chemical
potentials for the global charges.
\lfm
\item{2)} In the unbroken phase in front of the expanding bubble walls
there will be
 net densities $\rho_a$  for various  global charges with weak anomalies
resulting from charge transport of particles reflected from the bubbles.
Compute these densities $\rho_a$; it is convenient, although not
necessary, to choose a basis
where only one of these anomalous global charges is nonzero.
\lfm
\item{3)} Compute the chemical potentials $\mu_a$ for the anomalous global
charges from eq.~\numbdens. In an orthogonal basis the result is \eqn\chemsol{
\mu_a={6 \rho_a\over T^2\sum_jk_j (q_j^{a})^2}\ }
where $j$ is summed over all particle species.
\lfm
\item{4)} Compute the baryon production rate from eq.~\barcreate.
\lfm
We can now redo the baryogenesis calculation of ref.~\usthree, working in a
basis where all global charges are orthogonal to hypercharge, and taking
into account hypercharge screening through the chemical potential $\mu_Y$.
In this basis several global charges which are  orthogonal to hypercharge are
nonzero in the unbroken phase. Since the result for the baryon number
production is
independent of the chemical potential $\mu_Y$, and also of course  basis
independent, we get exactly the same result for $\dot\rho_B$, as
confirmed by explicit computation\foot{up to a factor of 3/2 which was an
algebraic error in the computation of the baryon violation rate in
ref.~\usthree. In that paper, eq.~(C.2) should read
$\mu_B=2\rho_Y/((1+2n)T^2)$, and eq.~(2.11) should be
$\dot\rho_B=-(3\Gamma_B/T)(\partial F/\partial B).$}. We conclude that the
charge transport mechanism is unaffected by Debye screening and dominates
other baryon production mechanisms if the phase boundary is thin,  of
width $\ltap \CO(1/T)$.

 \bigbreak\bigskip\bigskip\centerline{\bf
Acknowledgments}\nobreak
We thank P. Diamond for useful conversations on plasma screening.   A. C. was
supported in part by DOE contracts \#DE-AC02-89ER40509,
\#DE-FG02-91ER40676,  by the Texas National
Research Laboratory Commission grant \#RGFY91B6,
and NSF contract \#PHY-9057173;
A. N. and D. K. were
supported in part by DOE contract \#DE-FG03-90ER40546 and by   fellowships
from the Alfred P. Sloan Foundation. D.K. was supported in part by NSF
contract PHY-9057135. A.N. was supported in part by an SSC Fellowship from
the Texas National Research Laboratory Commission.
 \bigbreak\bigskip\bigskip

\listrefs
\bye